\documentclass[aps,showpacs,twocolumn,prd,superscriptaddress]{revtex4}
\usepackage{epsfig,amsmath}

\begin{document}

\title{Relativistic static thin dust disks with an inner edge: \\
An infinite family of new exact solutions}

\author{Guillermo A. Gonz\'alez}\email{guillego@uis.edu.co}
\affiliation{Escuela de F\'{\i}sica, Universidad Industrial de Santander, A. A.
678, Bucaramanga, Colombia}
\affiliation{Departamento de F\'{\i}sica Te\'orica, Universidad del Pa\'is
Vasco, 48080 Bilbao, Spain}

\author{Antonio C. Guti\'errez-Pi\~{n}eres}\email{agutierrez@ciencias.uis.edu.co}
\affiliation{Escuela de F\'{\i}sica, Universidad Industrial de Santander, A. A.
678, Bucaramanga, Colombia}

\author{Viviana M. Vi\~{n}a-Cervantes}\email{vivo1225@gmail.com}
\affiliation{Escuela de F\'{\i}sica, Universidad Industrial de Santander, A. A.
678, Bucaramanga, Colombia}

\pacs{04.20.-q, 04.20.Jb, 04.40.-b}

\begin{abstract}
An infinite family of new exact solutions of the Einstein vacuum equations for
static and axially symmetric spacetimes is presented. All the metric functions
of the solutions are explicitly computed and the obtained expressions are simply
written in terms of oblate spheroidal coordinates. Furthermore, the solutions
are asymptotically flat and regular everywhere, as it is shown by computing all
the curvature scalars. These solutions describe an infinite family of thin dust
disks with a central inner edge, whose energy densities are everywhere positive
and well behaved, in such a way that  their energy-momentum tensor are in fully
agreement with all the energy conditions. Now, although the disks are of
infinite extension, all of them have finite mass. The superposition of the first
member of this family with a Schwarzschild black hole was presented previously
[G. A. Gonz\'alez and A. C. Guti\'errez-Pi\~neres, arXiv: 0811.3002v1 (2008)],
whereas that in a subsequent paper a detailed analysis of the corresponding
superposition for the full family will be presented.
\end{abstract}

\maketitle

\section{\label{sec:intro}Introduction}

The observational data supporting the existence of black holes at the nucleus of
some galaxies is today so abundant, with the strongest dynamical evidence coming
from the center of the Milky Way, that there is no doubt about the relevance of
the study of binary systems composed by a thin disk surrounding a central black
hole (see \cite{CMS,BEG} for recent reviews on the observational evidence).
Accordingly, a lot of work has been developed in the last years in order to
obtain a better understanding of the different aspects involved in the dynamics
of these systems. Now, due to the presence of a black hole, the gravitational
fields involved are so strong that the proper theoretical framework to
analytically study this configurations is provided by the general theory of
relativity. Therefore, a strong effort has been dedicated to the obtention of
exact solutions of Einstein equations corresponding to thin disklike sources
with a central black hole (see \cite{SEM1,KHS} for thoroughly surveys on the
subject).

Stationary and axially symmetric solutions of the Einstein equations are of
obvious astrophysical importance, as they describe the exterior of equilibrium
configurations of bodies in rotation. At the same time, such spacetimes are the
best choice to attempt to describe the gravitational fields of disks around
black holes in an exact analytical manner. So, through the years, several
examples of solutions corresponding to black holes or to thin disklike sources
has been obtained by many different techniques. However, due to the nonlinear
character of the Einstein equations, solutions corresponding to the
superposition of black holes and thin disks are not so easy to obtain and so,
until now, exact stationary solutions have not been obtained.

On the other hand, if we only consider static configurations, the line element
is characterized only by two metric functions. So, in the vacuum case, the
Einstein equations implies that one of the metric functions satisfies the
Laplace equation whereas that the other one can be obtained by quadratures.
Furthermore, as the sources are infinitesimally thin disks, the matter only
enters in the form of boundary conditions for the vacuum equations. Therefore,
as a consequence of the linearity of the Laplace equation, solutions
corresponding to the superposition of thin disks and black holes can be, in
principle, easily obtained.

However, if we consider thin disks that extend up to the event horizon, the
matter located near the black hole will moves with superluminal velocities, as
was shown by Lemos and Letelier \cite{LL1, LL2, LL3}. So, in order to prevent
the appearance of tachyonic matter, the thin disks must have an inner edge with
a radius larger than the photonic radius of the black hole. Then, the boundary
value problem for the Laplace equation is mathematically more complicated and
thus only very few exact solutions had been obtained. These kind of solutions
were first studied by Lemos and Letelier \cite{LL2} by making a Kelvin
transformation in order to invert the Morgan and Morgan \cite{MM1} family of
finite thin disks. Now, although the second metric function of this solution can
not be analytically obtained, their main properties were extensively analyzed in
a series of papers by Semer\'ak, \u{Z}\'a\u{c}ek and Zellerin \cite{SZZ1, SZZ2,
SZ1, SZ2, SEM2, ZS1, SEM3}, by using numerical computation when was needed.

Besides the Lemos and Letelier inverted disks, only two other solutions for
static thin disks with an inner edge have been obtained, a first one with
inverted isochrone disks \cite{K1} and a second one for disks with a power-law
density \cite{SEM4}. Also, a stationary superposition was obtained by Zellerin
and Semer\'ak \cite{ZS2} by using the Belinskii-Zakahrov inverse-scattering
method, but the analysis of its properties is complicated by the fact that the
metric functions can not be analytically computed. Furthermore, this solution
involves an unphysical supporting surface between the black-hole horizon and the
disk \cite{SEM5}. Finally, a general class of stationary solutions was presented
by Klein \cite{K2}, by using the Riemann-surface techniques, with which
physically acceptable black hole disk systems can in principle be found.

Now, a common feature of all the above mentioned solutions is that their metric
functions can not be fully analytically computed. Thus then, the analysis of
their physical and mathematical properties is very complicated. Moreover, almost
all of these solutions do present singularities at the inner edge of the disk,
perhaps the only exemption being the class of solutions presented by Klein
\cite{K2}. Now then, the ubiquitous presence of this singularity at almost all
the obtained solutions has been considered by some authors as tightly connected
with the unphysical infinite thinness of the source. However, as we will show at
this paper, if the corresponding boundary value problem is properly solved, it
is possible to obtain a whole family of singularity free solutions.

In this paper we present an infinite family of new exact solutions for static
thin dust disks with a central inner edge. These solutions describe disks whose
energy densities are everywhere positive and well behaved, in such a way that
their energy-momentum tensor are in fully agreement with all the energy
conditions. Now, although the disks are of infinite extension, all of them have
finite mass. Furthermore, the solutions are asymptotically flat and regular
everywhere, as it is shown by computing all the curvature scalars. The
superposition of the first member of this family with a Schwarzschild black hole
was presented previously \cite{GG1}, whereas that in a subsequent paper a
detailed analysis of the corresponding superposition for the full family will be
presented.

The paper is organized as follows. First, in Sec. \ref{sec:eqs}, we present the
formulation of the Einstein equations for static axially symmetric spacetimes
with an infinitesimally thin disk as source. We also present the proper boundary
conditions and their relationship with the physical quantities characterizing
the sources. Then, in Sec. \ref{sec:sols}, we introduce oblate spheroidal
coordinates with the ranges choosen in such a way that they be naturally adapted
to the geometry of a thin disk with a central inner edge. The Einstein equations
are then solved and the metric functions of whole the family of solutions are
explicitly computed and the obtained expressions are simply written in terms of
the oblate spheroidal coordinates. The behavior of the solutions is then
analysed in Sec. \ref{sec:beh}. The analysis is made by studying the behavior of
all the curvature invariants for the full family as well as the behavior of the
corresponding energy densities and azimuthal pressures. Also we analyze the mass
densities of the disks and the finiteness of their corresponding total mass.
Finally, in Sec. \ref{sec:con}, we summarize the results.

\section{The Einstein equations with thin disklike sources}\label{sec:eqs}

In order to formulate the Einstein equations for static axially symmetric
spacetimes with an infinitesimally thin disk as source, first we introduce
coordinates $x^a = (t,\varphi,r,z)$ in which the metric tensor only depends on
$r$ and $z$. We assume that these coordinates are quasicylindrical in the sense
that the coordinate $r$ vanishes on the axis of symmetry and, for fixed $z$,
increases monotonically to infinity, while the coordinate $z$, for fixed $r$,
increases monotonically at the interval $(-\infty,\infty)$. The azimuthal angle
$\varphi$ ranges at the interval $[0,2\pi)$, as usual \cite{MM2}. Also we assume
that there exists at the spacetime an infinitesimally thin disk, located at the
hypersurface $z = 0$, in such a way that the components of the metric tensor are
symmetrical functions of $z$ and that their first $z$-derivatives have a finite
discontinuity at $z = 0$.

Accordingly with the above considerations,
\begin{equation}
g_{ab} (r,-z) = g_{ab} (r,z), \label{eq:simg}
\end{equation}
in such a way that, for $z \neq 0$,
\begin{equation}
g_{ab,z} (r,-z) = - g_{ab,z} (r,z). \label{eq:simgz}
\end{equation}
Thus then, the metric tensor is continuous at $z = 0$,
\begin{equation}
[g_{ab}] = g_{ab}|_{_{z = 0^+}} - g_{ab}|_{_{z = 0^-}} = 0 ,
\end{equation}
whereas that the discontinuities in the derivatives of the metric tensor can be
written as
\begin{equation}
\gamma_{ab} =  [{g_{ab,z}}] = 2 g_{ab,z}|_{_{z = 0^+}},
\end{equation}
where the reflection symmetry with respect to $z = 0$ has been used. So, by
using the distributional approach \cite{PH, LICH, TAUB}, we can write the metric
tensor as
\begin{equation}
g_{ab} = g^+_{ab} \theta (z) + g^-_{ab} \{ 1 - \theta (z) \},
\end{equation}
in such a way that the Ricci tensor can be written as
\begin{equation}
R_{ab} = R^+_{ab} \theta(z) + R^-_{ab} \{ 1 - \theta (z) \} + H_{ab} \delta(z),
\label{eq:ricdis}
\end{equation}
where $\theta(z)$ and $\delta (z)$ are, respectively, the Heaveside and Dirac
distributions with support on $z = 0$. Here $g^\pm_{ab}$ and $R^\pm_{ab}$ are
the metric tensors and the Ricci tensors of the $z \geq 0$ and $z \leq 0$
regions, respectively, whereas that
\begin{eqnarray}
H_{ab} = \frac{1}{2} \{ \gamma^z_a \delta^z_b  + \gamma^z_b \delta^z_a
-\gamma^c_c \delta^z_a \delta^z_b - g^{zz} \gamma_{ab} \},
\end{eqnarray}
where all the quantities are evaluated at $z = 0^+$.

Then, in agreement with (\ref{eq:ricdis}), the energy-momentum tensor must be
expressed as 
\begin{equation}
T_{ab} = T^+_{ab} \theta(z) + T^-_{ab} \{ 1 - \theta(z) \} + Q_{ab} \delta(z),
\label{eq:emtot}
\end{equation}
where $T^\pm_{ab}$ are the energy-momentum tensors for the $z \geq 0$ and $z
\leq 0$ regions, respectively, and $Q_{ab}$ gives the part of the
energy-momentum tensor corresponding to the disk source.  Accordingly, the
Einstein equations, in geometrized units such that $c = 8 \pi G = 1$, are
equivalent to the system
\begin{eqnarray}
R^\pm_{ab} - \frac{1}{2} g_{ab} R^\pm &=& T^\pm_{ab}, \label{eq:einspm} \\
&	&	\nonumber \\
H_{ab} - \frac{1}{2} g_{ab} H &=& Q_{ab}, \label{eq:einsdis}
\end{eqnarray}
where $H = g^{ab} H_{ab}$ and, again, all the quantities are evaluated at $z =
0^+$. Now then, when the thin disk is the only source of the gravitational
field, so that $T^\pm_{ab} = 0$, equation (\ref{eq:einspm}) reduces to the
Einstein vacuum equations
\begin{equation}
R^\pm_{ab} = 0, \label{eq:einsvac}
\end{equation}
for the $z \geq 0$ and $z \leq 0$ regions, respectively. Thus, in order to
obtain solutions wtih a thin disk as source, we must solve the system
(\ref{eq:einsvac}) by using in equation (\ref{eq:einsdis}), as boundary
conditions, the values of $Q_{ab}$ that describe properly the matter content of
the disk.

Now, in order to obtain explicit forms for the vacuum Einstein equations and the
boundary conditions, we will take the metric tensor as given by the Weyl line
element, written as \cite{KSMH}
\begin{eqnarray}
ds^2 = - \ e^{2 \Phi} dt^2 + e^{- 2 \Phi}[r^2  d\varphi^2 + e^{2\Lambda} (dr^2 +
dz^2)], \label{eq:met}
\end{eqnarray} 
where $\Phi$ and $\Lambda$ are continuous functions of $r$ and $z$. Furthermore,
we will assume that $\Phi$ and $\Lambda$ are even functions of $z$,
\begin{subequations}\begin{eqnarray}
\Phi(r,-z) &=& \ \ \Phi(r,z), \label{eq:con1a} \\
&&	\nonumber	\\
\Lambda(r,-z) &=& - \Lambda(r,z), \label{eq:con1b}
\end{eqnarray}\label{eq:con1}\end{subequations}
in such a way that their first $z$-derivatives are odd functions of $z$,
\begin{subequations}\begin{eqnarray}
\Phi_{,z}(r,-z) &=& - \Phi_{,z}(r,z), \label{eq:con2a} \\
&&	\nonumber	\\
\Lambda_{,z}(r,-z) &=& - \Lambda_{,z}(r,z), \label{eq:con2b}
\end{eqnarray}\label{eq:con2}\end{subequations}
which we shall require that do not vanish at $z = 0$. So, the vacuum Einstein
equations (\ref{eq:einsvac}) are equivalent to the system
\begin{subequations}\begin{eqnarray}
&&(r \Phi_{,r})_{,r} + (r \Phi_{,z})_{,z} = 0, \label{eq:ein1} \\
&&	\nonumber	\\
&&\Lambda_{,r} = r (\Phi^2_{,r} - \Phi^2_{,z} ), \label{eq:ein2} \\
&&	\nonumber	\\
&&\Lambda_{,z} = 2 r \Phi_{,r} \Phi_{,z}, \label{eq:ein3}
\end{eqnarray}\label{eq:ein}\end{subequations} 
where (\ref{eq:ein1}) is the usual Laplace equation for an axially symmetric
source in cylindrical coordinates, whereas that the integrability condition for
the overdetermined system (\ref{eq:ein2}) - (\ref{eq:ein3}) is granted when
$\Phi$ is a solution of (\ref{eq:ein1}), in such a way that $\Lambda$ can be
obtained by quadratures given a solution for $\Phi$.

On the other hand, equation (\ref{eq:einsdis}) implies that the boundary
conditions reduce to
\begin{subequations}\begin{eqnarray}
2 e^{2(\Phi - \Lambda)} \left[ \Lambda_{,z} - 2 \Phi_{,z} \right] &=&
Q_{t}^{t},
\label{eq:emt1}  \\
&	&	\nonumber	\\
2 e^{2(\Phi - \Lambda)} \Lambda_{,z} &=& Q_{\varphi}^{\varphi} \label{eq:emt2},
\end{eqnarray}\label{eq:emt}\end{subequations}
where all the quantities are evaluated at $z = 0^+$, and that $Q_{ab}$ must have
only two nonzero components. So, by using the orthonormal tetrad
\begin{subequations}\begin{eqnarray}
V^a &=& e^{- \Phi} \ \delta^a_{t} ,	\\
	&	&	\nonumber	\\
W^a &=& {e^\Phi} \ \delta^a_{\varphi} / r,	\\
	&	&	\nonumber	\\
X^a &=& e^{\Phi - \Lambda} \ \delta^a_{r} ,	\\
	&	&	\nonumber	\\
Y^a &=& e^{\Phi - \Lambda} \ \delta^a_{z} ,
\end{eqnarray}\label{eq:tetrad}\end{subequations}
we can write the surface energy-momentum tensor $Q_{ab}$ in the canonical form
\begin{equation}
Q_{ab} \ = \ \epsilon V_a V_b + p W_a W_b, \label{eq:emtdia}
\end{equation}
where $\epsilon$ and $p$ are, respectively, the energy density and the azimuthal
pressure of the disk. In terms of these quantities, the boundary conditions can
be written as
\begin{eqnarray}
2 e^{2(\Phi - \Lambda)} \left[ 2 \Phi_{,z} - \Lambda_{,z} \right] &=& \epsilon,
\label{eq:energy} \\
&&	\nonumber	\\
2 e^{2(\Phi - \Lambda)} \Lambda_{,z} &=& p. \label{eq:press}
\end{eqnarray}
Finally, by using (\ref{eq:ein3}), we can cast these conditions as
\begin{eqnarray}
4 e^{2(\Phi - \Lambda)} \left[1 - r \Phi_{,r} \right] \Phi_{,z} &=& \epsilon,
\label{eq:energy2} \\
&&	\nonumber	\\
4 e^{2(\Phi - \Lambda)} r \Phi_{,r} \Phi_{,z} &=& p, \label{eq:press2}
\end{eqnarray}
where, as before, all the quantities are evaluated at $z = 0^+$.

As we can see from the above expressions, the more general energy-momentum
tensor that is compatible with the line element (\ref{eq:met}) and the boundary
conditions (\ref{eq:einsdis}), corresponds to a thin disklike source that only
have a nonzero energy density and a nonzero azimuthal pressure. In agreement
with this, instead of give specific prescriptions for the energy density and the
azimuthal pressure, the Einstein equations will be solved only by requiring that
these two quantities will be different from zero at the surface of a disk with
an inner edge. Then, after a given solution be obtained, it can be used in order
to obtain, from the boundary conditions, the corresponding expressions for the
energy density and the azimuthal pressure. Therefore, the solution will
correspond to the more general static thin disk with an inner edge that can be
obtained by exactly solving the Einstein equations.

Accordingly, in order to obtain a solution representing a thin disk located in
the hypersurface $z = 0$, with a circular central inner edge of radius $a$, we
only need to imposse that
\begin{eqnarray}
\Phi_{,z}(r,0^{+}) &=& \left\{ \begin{array}{cl}
0 &; \;\; 0 \leq r \leq a, \\
	&	\\
f(r) & ; \;\;r \geq a, \\
\end{array}\label{eq:con3}\right. 
\end{eqnarray}
with $f(r)$ any arbitrary function. Then, only after we find the more general
solution, we will impose additional conditions in order to have a physically
razonable behavior. So, in order to have an asymptotically flat spacetime, we
will require that
\begin{subequations}\begin{eqnarray}
\lim_{{\rm R} \to \infty} \Phi (r,z) &=& 0, \label{eq:limphi} \\
&&	\nonumber	\\
\lim_{{\rm R} \to \infty} \Lambda (r,z) &=& 0, \label{eq:limlam}
\end{eqnarray}\label{eq:lims}\end{subequations}
where ${\rm R}^2 = r^2 + z^2$. Also, in order to have regularity at the symmetry
axis, we will require that
\begin{subequations}\begin{eqnarray}
\Phi(0,z) &<& \infty, \label{eq:phi0} \\
&&	\nonumber \\
\Lambda(0,z) &<& \infty. \label{eq:lam0}
\end{eqnarray}\label{eq:regul}\end{subequations}
We also will require that
\begin{subequations}\begin{eqnarray}
f(r) \geq 0,&& \label{eq:enpos1} \\
&&	\nonumber	\\
0 \leq r \Phi_{,r} \leq 1,&& \label{eq:enpos2}
\end{eqnarray}\label{eq:enpos}\end{subequations}
in order that the energy density and the azimuthal pressure be positive
everywhere.

Finally, we need to check the finiteness of the total mass of the disks. So, in
order to do this, first we take the mass density $\mu$ of the disks as defined
by
\begin{equation}
\frac{\mu}{2} = ( Q_{ab} - \frac{1}{2} g_{ab} Q ) V^a V^b,
\end{equation}
where $Q = g^{ab} Q_{ab}$ and, as before, all the quantities are evaluated at $z
= 0^+$. Accordingly, by using (\ref{eq:emt}) and (\ref{eq:emtdia}), we have that
the mass density reduces to
\begin{equation}
\mu = \epsilon + p. \label{eq:masden}
\end{equation}
On the other hand, the total mass of the disks is given by
\begin{equation}
M = \int \mu d\Sigma = \int_{0}^{2\pi} \int_{a}^{\infty} \mu e^{\Lambda -
2\Phi} r dr d\varphi,
\end{equation}
where $d\Sigma = e^{\Lambda - 2\Phi} r dr d\varphi$ is the area element on the
disk surface. So, by using (\ref{eq:masden}), we can write the total mass as
\begin{equation}
M = 2 \pi \int_{a}^{\infty} (\epsilon + p) e^{\Lambda - 2\Phi} r dr,
\label{eq:mass}
\end{equation}
where the integration on $\varphi$ has been made.

\section{Solution of the Einstein equations}\label{sec:sols}

In order to solve the Einstein vacuum equations, first we must solve the
boundary value problem for $\Phi$. However, due to the nature of the boundary
conditions (\ref{eq:con2}), it is convenient to look for a different coordinate
system that be naturally adapted to the geometry of the desired source.
Accordingly, we introduce the oblate spheroidal coordinates as defined trough
the relations
\begin{subequations}\begin{eqnarray}
r^2 &=& a^2 (1 + x^2) (1- y^2), \label{eq:rxy} \\
&&	\nonumber	\\
z &=& a x y, \label{eq:zxy}
\end{eqnarray}\end{subequations}
where $-\infty < x < \infty$ and $0 \leq y \leq 1$. So, when $x = 0$ we have
that $z = 0$ and $0 \leq r \leq a$, whereas that when $y = 0$ we have that $z =
0$ and $r \geq a$. Furthermore, as $x$ changes sign on crossing the surface $y =
0$, but do not changes in absolute value, this coordinate has a finite
discontinuity when $y = 0$. Thus then, an even function of $x$ is a continuous
fuction everywhere but has a discontinuous normal derivative at $y = 0$. On the
other hand, $y$ is continuous everywhere. Therefore, the surface $y = 0$
describes a thin disk with an inner edge of radius $a$, whereas that the surface
$x = 0$ describes the vacuum hole inside this edge.

In the oblate spheroidal coordinates, the Weyl line element (\ref{eq:met}) can
be rewritten as
\begin{eqnarray}
ds^2 &=&- \ e^{2 \Phi} dt^2 + a^2 (1 + x^2) (1 - y^2) e^{- 2 \Phi} d\varphi^2 
\quad \quad \quad \nonumber \\
&&+ \ a^2 (x^2 + y^2) e^{2(\Lambda - \Phi)} \left[ \frac{dx^2}{1 + x^2} +
\frac{dy^2}{1 - y^2} \right], \label{eq:metob}
\end{eqnarray}
in such a way that the Einstein vacuum equations reduce to
\begin{eqnarray}
[ (1 + x^2) \Phi_{,x} ]_{,x} + [ (1 - y^2) \Phi_{,y} ]_{,y} = 0,
\label{eq:lapoblate}
\end{eqnarray}
the Laplace equation in oblate spheroidal coordinates, and the overdetermined
system
\begin{eqnarray}
\Lambda_{,x} &=& (1 - y^2) \left[ x(1 + x^2)\Phi_{,x}^2 - x(1 - y^2)\Phi_{,y}^2
\right. \nonumber \\
&&\left. - \ 2y(1 + x^2)\Phi_{,x}\Phi_{,y} \right] / (x^2 + y^2), \label{eq:lax}
\\
&&	\nonumber \\
\Lambda_{,y} &=& (1 + x^2) \left[ y(1 + x^2)\Phi_{,x}^2 - y(1 - y^2)\Phi_{,y}^2
\right. \nonumber \\
&&\left. + \ 2x(1 - y^2)\Phi_{,x}\Phi_{,y } \right] / (x^2 + y^2),
\label{eq:lay}
\end{eqnarray}
whose integrability again is granted by equation (\ref{eq:lapoblate}).

On the other hand, by using (\ref{eq:rxy}) and (\ref{eq:zxy}), is easy to see
that 
\begin{equation}
\Phi_{,z}(r,0) = \left\{ \begin{array}{ll}
\Phi_{,x} (0,y) / a y &; \;\; 0 \leq r \leq a, \\
	&	\\
\Phi_{,y} (x,0) / a x & ; \;\;r \geq a. \\
\end{array}\right. \label{eq:phiz}
\end{equation}
Accordingly, as the reflection symmetry of the solutions implies that
\begin{subequations}\begin{eqnarray}
\Phi(-x,y) &=& \ \Phi(x,y), \label{eq:conpar1} \\
&&	\nonumber	\\
\Phi_{,x}(-x,y) &=& - \Phi_{,x}(x,y), \label{eq:conpar2}
\end{eqnarray}\label{eq:conpar}\end{subequations}
the conditions (\ref{eq:con3}) are equivalent to
\begin{subequations}\begin{eqnarray}
\Phi_{,x}(0,y) &=& 0, \label{eq:conob1} \\
&&	\nonumber	\\
\Phi_{,y}(x,0) &=& F(x); \quad x \geq 0, \label{eq:conob2}
\end{eqnarray}\label{eq:conob}\end{subequations}
with $F(x)$ any arbitrary function. The general solution of equation
(\ref{eq:lapoblate}) with these boundary conditions is given by \cite{BAT}
\begin{equation}
\Phi(x,y) = \sum_{n = 0}^{\infty} [ A_{2n} P_{2n}(y) + B_{2n} Q_{2n}(y) ]
p_{2n}(x), \label{eq:solaplac}
\end{equation}
where $A_{2n}$ and $B_{2n}$ are constants, $P_{2n}(y)$ and $Q_{2n}(y)$ are the
Legendre polynomials and the Legendre functions of the second kind,
respectively, and $p_{2n}(x) = i^{-2n} P_{2n}(ix)$. Therefore, all the solutions
of the Einstein vacuum equations for static spacetimes with any axially
symmetric source as the considered here, are obtained by taking for the metric
function $\Phi(x,y)$ any particular choice of the above general solution, or
expressions obtained from these solutions by means of linear operations. 

Now, in terms of the oblate spheroidal coordinates, condition (\ref{eq:limphi})
is written as
\begin{equation}
\lim_{x \to \infty} \Phi(x,y) = 0,
\end{equation}
whereas condition (\ref{eq:phi0}) is written as
\begin{equation}
\Phi(x,1) < \infty.
\end{equation}
So, due to the behavior of the Legendre functions, it is clear that it is not
possible to fulfill the physical conditions (\ref{eq:lims}) and (\ref{eq:regul})
with any particular choice of the general solution (\ref{eq:solaplac}). However,
by considering only the first term of the series,
\begin{equation}
\Phi_{0}(x,y) = A_0 + B_0 Q_0 (y),
\end{equation}
we obtain a solution that is regular for all $y \neq 1$. Then, if we take $A_0 =
0$, this solution can be written as
\begin{eqnarray}
\Phi_{0}(x,y) = \frac{\alpha}{2} \ln \left[\frac{1 + y}{1 -y}\right]
\label{eq:phidef},
\end{eqnarray}
where $\alpha$ is an arbitrary constant, in such a way that a direct integration
of (\ref{eq:lax})-(\ref{eq:lay}) gives
\begin{equation}
\Lambda_{0}(x,y) =  \frac{\alpha^2}{2} \ln \left[\frac{1 - y^2}{x^2 +
y^2}\right],
\end{equation}
where the integration constant has been taken as zero. As we can see, this
solution is not asymptotically flat neither regular at the symmetry axis.
Nevertheless, by using (\ref{eq:energy}) and (\ref{eq:press}), we obtain for the
energy density and the azimuthal pressure the expressions
\begin{eqnarray}
\epsilon &=& ({4\alpha}/{a}) x^{2\alpha^2 -1}, \label{eq:density} \\
&&	\nonumber	\\
p &=& 0, \label{eq:presure}
\end{eqnarray}
in such a way that, if $\alpha > 0$, the disk satisfy all the energy contions
\cite{HE}. However, for any value of $\alpha^2 \neq \frac{1}{2}$, the energy
density increases without limit, may be at infinity or at the inner edge of the
disk, whereas that for $\alpha^2 = \frac{1}{2}$ the energy density is everywhere
constant, so that, in any case, the total mass of the disk will be infinite.

On the other hand, although the previous solution has not a physically
acceptable behavior, we can use it as the starting point to generate new and
well behaved solutions. In order to do this, we consider the oblate spheroidal
coordinates not only as functions of the cylindrical coordinates $(r,z)$, but
also as parametrically dependents of the radius $a$,
\begin{subequations}\begin{eqnarray}
x &=& x (r, z; a) , \\
&&	\nonumber \\
y &=& y (r, z; a) . 
\end{eqnarray}\label{eq:xya}\end{subequations}
Accordingly, by considering also the metric function $\Phi$ as dependent of $a$,
\begin{eqnarray}
\Phi &=& \Phi (r,z;a),
\end{eqnarray}
we can obtain a family of new solutions by applying the linear operation
\begin{equation}
\Phi_{n+1}(r,z;a) =  \frac{\partial \Phi_{n}(r,z;a)}{\partial a},
\label{eq:algorithm}
\end{equation}
where $n$ is an integer, $n \geq 0$.

Thus then, by starting with the ``seed solution'' $\Phi_{0}(x,y)$, by means of
the previous procedure it is generated a family of new solutions that can be
written in the simple form
\begin{equation}
\Phi_n (r,z;a) = \Phi_n (x,y) = \frac{\alpha y F_{n}(x,y)}{a^n (x^2 +
y^2)^{2n-1}}, 
\label{eq:phin}
\end{equation}
for $n \geq 1$, where the $F_n (x,y)$ are polynomial functions, with highest
degree $4n-4$, of which only we present below the first three,
\begin{eqnarray*}
F_1 &=& 1, \\
&& \\
F_2 &=& x^4 + 3x^2 (1 - y^2) - y^2, \\
&&	\\
F_3 &=& 3 x^6 (3 - 5 y^2) + 5 x^4 (6 y^4 - 11 y^2 + 3) \\
&-& x^2 y^2 (3 y^4 - 31 y^2 + 30) - y^4 (y^2 - 3), 
\end{eqnarray*}
but all of them can be easily obtained by means of (\ref{eq:algorithm}). So, is
easy to see that
\begin{equation}
\lim_{x \to \infty} \Phi_n (x,y) = 0,
\end{equation}
and that
\begin{equation}
\Phi_n (x,1) < \infty,
\end{equation}
in fully agreement with conditions (\ref{eq:limphi}) and (\ref{eq:phi0}). 

Now, in order to obtain the correspondig metric functions $\Lambda_n (r,z;a)$, we
make the integration
\begin{equation}
\Lambda_n (r,z;a) = \Lambda_n (x,y) = \int_1^y \Lambda_{,y} (x,y) dy,
\label{eq:intlam}
\end{equation}
by taking $\Lambda_n (x,1) = 0$ in order to grant regularity at the axis. So, by
using (\ref{eq:phin}) in (\ref{eq:lay}), the obtained solutions can be written
in the simple form
\begin{equation}
\Lambda_n (x,y) = \frac{\alpha^2 (2n - 2)! (y^2 - 1) A_n (x,y)}{4^n a^{2n}
(x^2 + y^2)^{4n}}, \label{eq:lamn}
\end{equation}
for $n \geq 1$, where the $A_n (x,y)$ are polynomial functions, off highest
degree $8n-2$, of which we present here only the first three,
\begin{eqnarray*}
A_1 &=& x^4 (9 y^2 - 1) + 2 x^2 y^2 (y^2 + 3) + y^4 (y^2 - 1), \\
&& \\
A_2 &=& 2 x^{12} (9 y^2 - 1) - 4 x^{10} (51 y^4 - 41 y^2 + 2) \\
&+& x^8 (735 y^6 - 1241 y^4 + 419 y^2 - 9) -x^6y^2 (132 y^6 \\
&-& 1644 y^4 + 1604 y^2 - 252) + x^4 y^4 (84 y^6 - 384 y^4  \\
&+& 1266y^2 - 630)+ 4 x^2 y^6 (6 y^6 + 6 y^4 - 39 y^2 + 63) \\
&+& \ 3 y^8 (y^6 + y^4 + y^2 - 3), \\
&&	\\
A_3 &=& 3 x^{16} (1225  y^6 - 1275 y^4 + 315 y^2 - 9) \\
&-& 24 x^{14} (980 y^8 - 2095 y^6 + 1205 y^4 - 189 y^2 + 3) \\
&+& 2 x^{12} (24255 y^{10} - 89475 y^8 + 98472 y^6 - 36316 y^4 \\
&+& 3473 y^2 - 25) - 12 x^{10} y^2 (1835 y^{10} - 16665 y^{8} \\
&+& 34716 y^6 - 25292 y^4 + 6001 y^2 - 275) \\
&+& 6 x^8 y^4 (900 y^{10} - 11946 y^8 + 50563 y^6 - 69397 y^4 \\
&+& 33365 y^2 - 4125) + 8 x^6 y^6 (125 y^{10} +926 y^8 \\
&-& 9079 y^6 + 24639 y^4 - 22290 y^2 + 5775) \\
&+& 6 x^4 y^8 (55 y^{10} + 29 y^8 + 764 y^6 - 4808 y^4 + 8469 y^2 \\
&-& 4125) + 12 x^2 y^{10} (5 y^{10} + 5 y^8 + 80 y^4 - 301 y^2\\
&+& 275) + y^{12} (5 y^{10} + 5 y^8 + 5 y^6 + y^4 + 34 y^2 - 50),
\end{eqnarray*}
but all of them can be obtained as a result of compute the integral
(\ref{eq:intlam}).  Accordingly, we have that
\begin{equation}
\lim_{x \to \infty} \Lambda_n (x,y) = 0,
\end{equation}
and that
\begin{equation}
\Lambda_n (x,1) = 0,
\end{equation}
in fully agreement with conditions (\ref{eq:limlam}) and (\ref{eq:lam0}).

\section{Behavior of the Solutions}\label{sec:beh}

As we can see from the expressions at previous section, by using
(\ref{eq:algorithm}) was generated an infinite family of asymptotically flat
solutions, all of them with a regular behavior at the symmetry axis. On the
other hand, it seems to be that there is a singularity at the inner edge of the
disks, when $x = 0$ and $y = 0$. However, is easy to see that this apparent
singularity is only a coordinate singularity. Indeed, as it is well known
\cite{KSMH,WEI}, the invariant characterization of the curvature must be in
terms of 14 scalars constructed from the Riemann tensor, $R_{abcd}$, and the
metric tensor, $g_{ab}$. So, in order to determine the nature of the aparent
singularities, we must to compute all the curvature invariants for the family of
solutions.

Now, for any solution of the Einstein vacuum equations, the curvature invariants
are the 10 vanishing components of the Ricci tensor, $R_{ab} = 0$, plus the four
scalars
\begin{eqnarray*}
{\cal K}_I &=& R^{abcd} R_{abcd}, \\
	&	&	\\
{\cal K}_{II} &=& {R^{ab}}_{kl} R^{klcd} R_{abcd}, \\
	&	&	\nonumber	\\
{\cal K}_{III} &=& \frac{{\epsilon^{ab}}_{kl} R^{klcd} R_{abcd}}{\sqrt{- g}},
\\
	&	&	\\
{\cal K}_{IV} &=& \frac{{\epsilon^{ab}}_{kl} {R^{kl}}_{mn} R^{mncd}
R_{abcd}}{\sqrt{- g}},
\end{eqnarray*}
where $g = \det g_{ab}$ and $\epsilon^{abcd}$ is the Levi-Civita symbol.
Furthermore, for any Weyl solution the last two invariants vanishes identically,
so that we only need to compute ${\cal K}_I$ and ${\cal K}_{II}$.

By using the expressions (\ref{eq:phin}) and (\ref{eq:lamn}) for $\Phi_n
(x,y)$ and $\Lambda_n (x,y)$, we can cast the curvature scalars as
\begin{subequations}\begin{eqnarray}
{\cal K}_{In} &=& - \frac{16 \alpha^2 e^{4(\Phi_n - \Lambda_n)} N_{In} (x,y)
}{a^{6n+4} (x^2 + y^2)^{12n}},  \label{eq:inv1} \\
&&	\nonumber	\\
{\cal K}_{IIn} &=& \ \frac{48 \alpha^3 e^{6(\Phi_n - \Lambda_n)} N_{IIn} (x,y)
}{a^{8n+6} (x^2 + y^2)^{16n}},  \label{eq:inv2} 
\end{eqnarray}\label{eq:invs}\end{subequations}
where $N_{In} (x,y)$ and $N_{IIn} (x,y)$ are polynomial functions, of highest
degree $24n-6$ and $32n-9$, respectively, that we do not write here explicitly
due to their long size, but that all of them vanishes at the inner edge of the
disks,
\begin{equation}
N_{In} (0,0) = N_{IIn} (0,0) = 0.
\end{equation}
Morover, is easy to see that, in any neighborhood around $(0,0)$, the
difference between $\Phi_n (x,y)$ and $\Lambda_n (x,y)$ behaves as
\begin{equation}
\Phi_n - \Lambda_n \approx - \frac{\alpha^2}{a^{2n} (x^2 + y^2)^{2n}},
\end{equation}
in such a way that
\begin{subequations}\begin{eqnarray}
\lim_{(x,y) \to (0,0)} \frac{e^{4(\Phi_n - \Lambda_n)}}{(x^2 + y^2)^{12n}} &=&
0, \\
&&	\nonumber	\\
\lim_{(x,y) \to (0,0)} \frac{e^{6(\Phi_n - \Lambda_n)}}{(x^2 + y^2)^{16n}} &=&
0,
\end{eqnarray}\end{subequations}
and the limits exist, whatever be the path chosen to approach the point
$(0,0)$. Accordingly, we have that
\begin{subequations}\begin{eqnarray}
\lim_{(x,y) \to (0,0)} {\cal K}_{In} (x,y) &=& 0, \\
&&	\nonumber	\\
\lim_{(x,y) \to (0,0)} {\cal K}_{IIn} (x,y) &=& 0,
\end{eqnarray}\end{subequations}
and thus the curvature is regular at the inner edge of the disks.

Now, in order to analyze the physical behavior of the sources, we will compute
the energy density and the azimuthal pressure for this family of disks. So,  by
using (\ref{eq:rxy}) and (\ref{eq:zxy}), we can see that
\begin{equation}
\Phi_{,r}(r,0) = \left[ \frac{\sqrt{1 + x^2}}{a x} \right] \Phi_{,x} (x,0),
\qquad r \geq a,
\end{equation}
and, by using (\ref{eq:phin}), is easy to prove that
\begin{equation}
\Phi_{n,x} (x,0) = 0, \qquad n \geq 1.
\end{equation}
Then, form (\ref{eq:press2}), we can see that
\begin{equation}
p_n = 0.
\end{equation}
That is, all the disks of the family have zero azimuthal pressure.

On the other hand, by using equations (\ref{eq:energy2}), (\ref{eq:phiz}) and
(\ref{eq:phin}), the surface energy density  of the disks can be written as
\begin{equation}
\epsilon_n (x) = \frac{4 \alpha E_n (x)}{a^{n+1} x^{2n+1}}\exp \left\{ -
\frac{\alpha^2 (2n-2)! B_n (x)}{2^{2n-1} a^{2n} x^{4n}} \right\},
\label{eq:enern}
\end{equation}
where $x \geq 0$ and the $E_n(x)$ are positive definite polynomials of degree
$2k$, with $k = (n-1)/2$ for odd $n$ and $k = n/2$ for even $n$, of which we
only will write below the first three,
\begin{eqnarray*}
E_1 (x) &=& 1, \\
&&	\\
E_2 (x) &=& x^2 + 3, \\
&&	\\
E_3 (x) &=& 3 (x^2 + 5),
\end{eqnarray*}
whereas that the $B_n (x)$ are positive definite polynomials of degree $4k$,
with $k = (n-1)/2$ for odd $n$ and $k = n/2$ for even $n$, the first three of
them given by
\begin{eqnarray*}
B_1 (x) &=& 1, \\
&&	\\
B_2 (x) &=& 2 x^4 + 8 x^2 + 9, \\
&&	\\
B_3 (x) &=& 27 x^4 + 72 x^2 + 50.
\end{eqnarray*}

From the above expressions we can see that, by taking $\alpha > 0$, the energy
density of the disks will be everywhere positive,
\begin{equation}
\epsilon_n (x) \geq 0.
\end{equation}
So that, as the azimuthal pressure is zero, we have an infinite family of dust
disks that are in fully agreement with all the energy conditions. Also is easy
to see that, for any value of $n$, we have that
\begin{subequations}\begin{eqnarray}
\epsilon_n (0) &=& 0, \\
&&	\nonumber	\\
\lim_{x \to \infty} \epsilon_n (x) &=& 0.
\end{eqnarray}\end{subequations}
That is, the energy density of the disks is zero at their inner edge and
vanishes at infinite. Furthermore, as the azimuthal pressure is zero, the mass
density of the disks reduces to their energy density,
\begin{equation}
\mu_n (x) = \epsilon_n (x), \label{eq:mun}
\end{equation}
so that its behavior is the same as the energy density. 

\begin{figure}
\begin{center}
$$\begin{array}{c}
\epsfig{width=3in,file=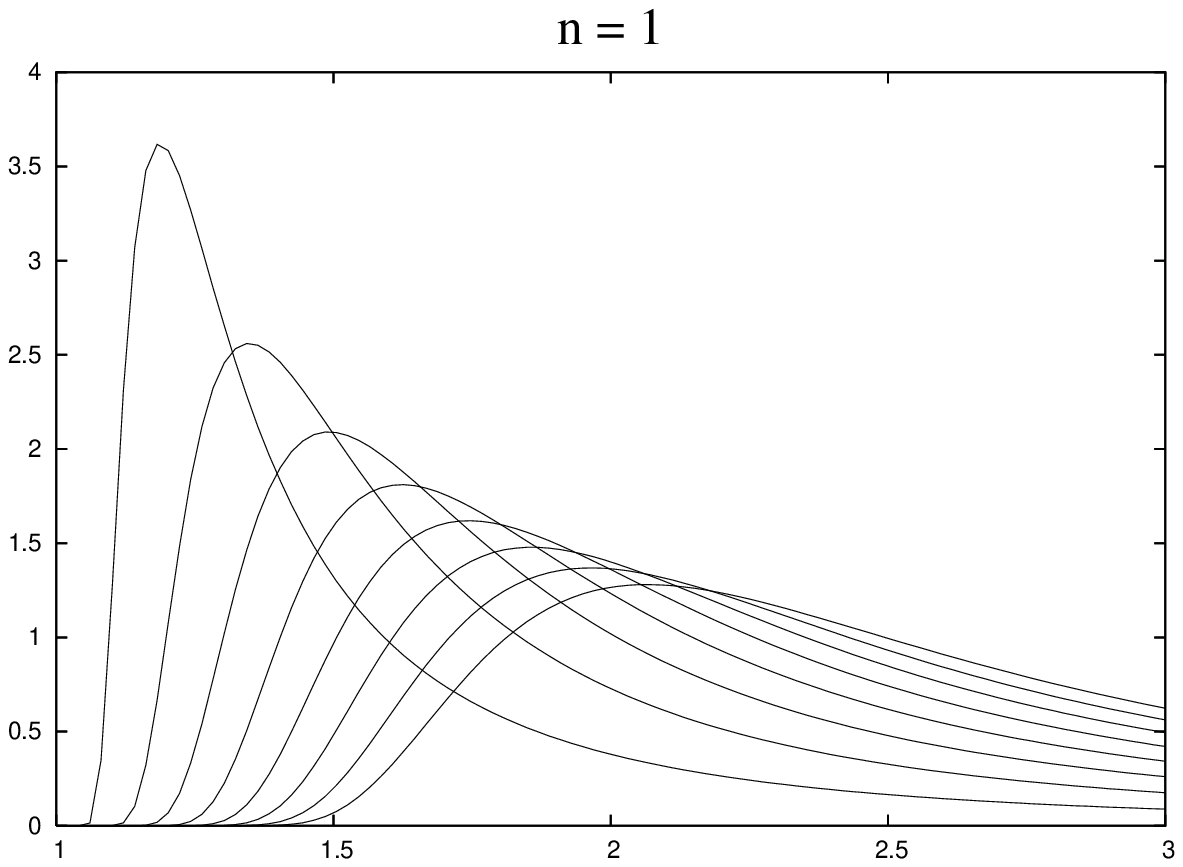} \\
\epsfig{width=3in,file=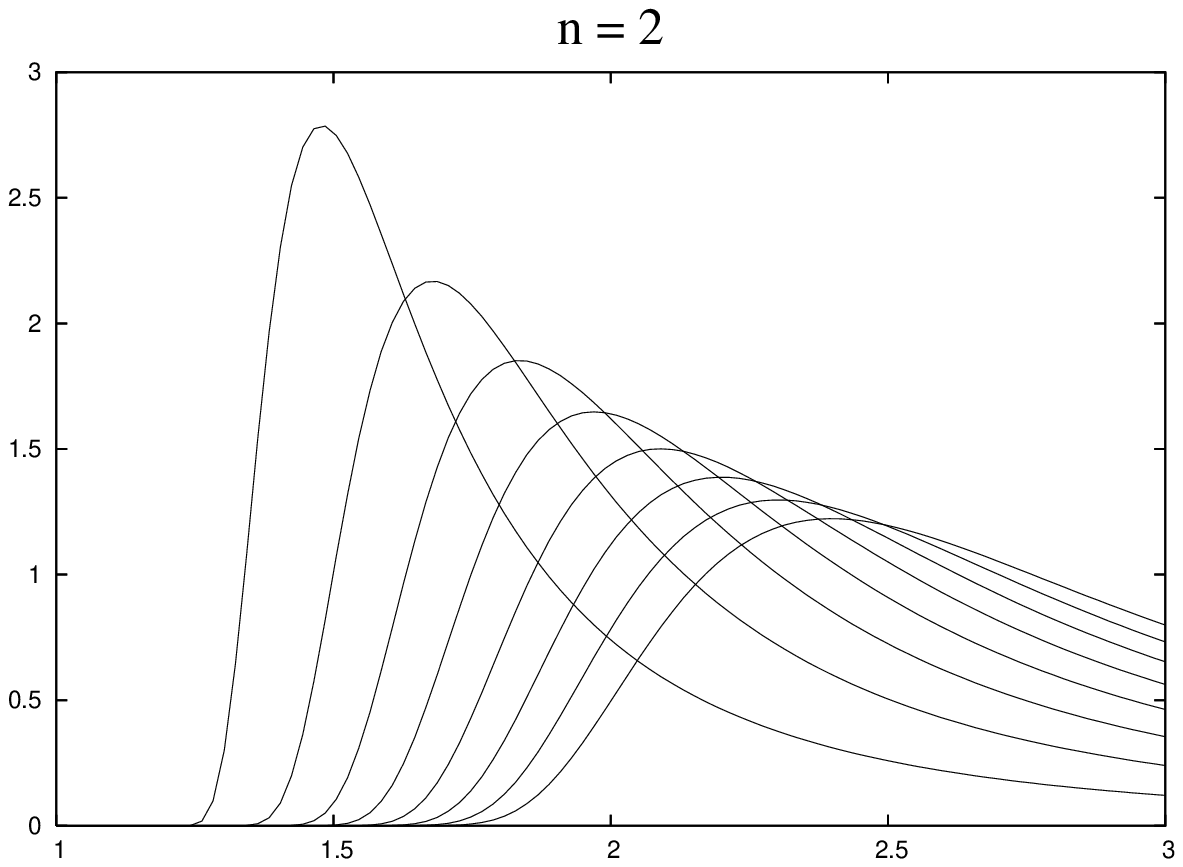} \\
\epsfig{width=3in,file=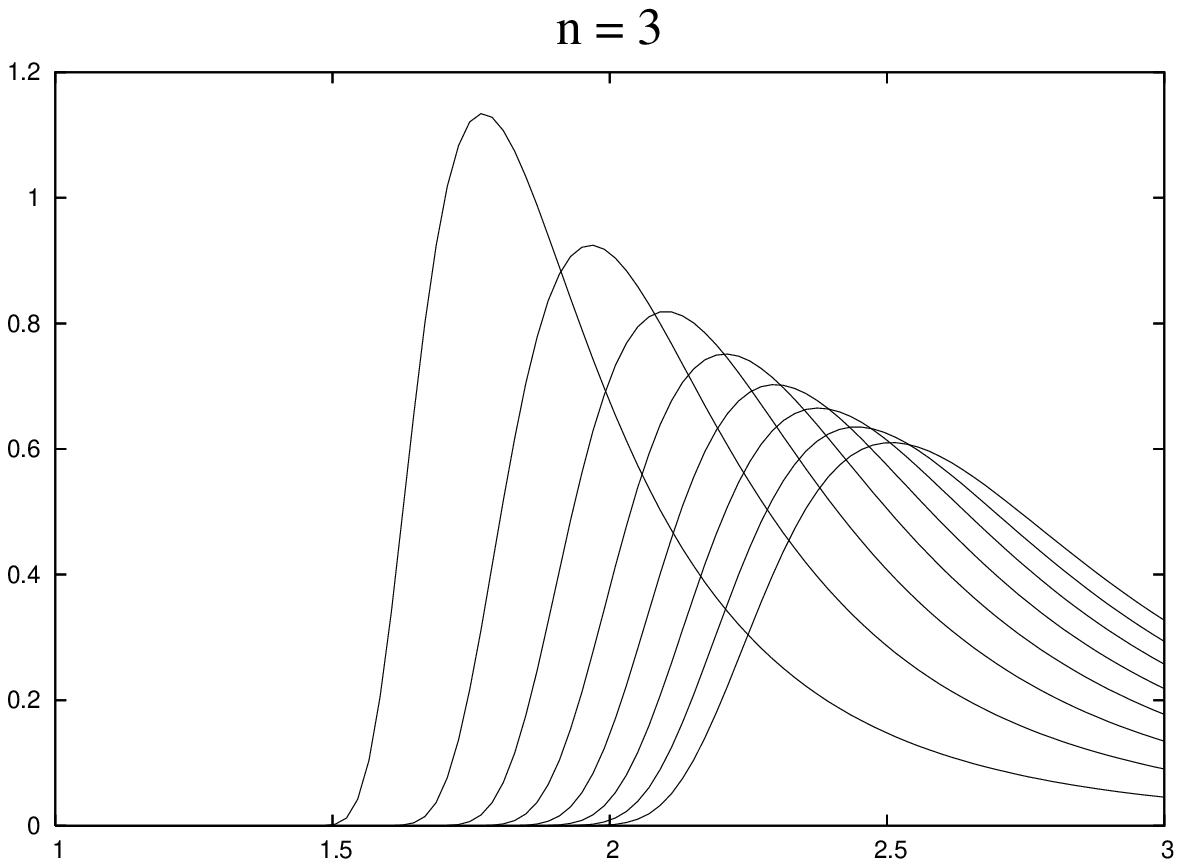} \\
\end{array}$$
\end{center}
\caption{\label{fig:energy}Surface energy density ${\tilde \epsilon}_n$ as a
function of ${\tilde r}$ for the first three disks of the family, with ${\tilde
\alpha}_n = 0.5$, $1$, $1.5$, $2$, $2.5$,  $3$, $3.5$ and $4$. For each value of
$n$, the first curve on left corresponds to ${\tilde \alpha} = 0.5$, whereas
that the last curve on right corresponds to ${\tilde \alpha} = 4$.}
\end{figure}

Now, in order to show the behavior of the energy densities, we plot the
dimensionless surface energy densities ${\tilde \epsilon}_n = a \epsilon_n$ as
functions of the dimensionless radial coordinate ${\tilde r} = r/a$. So, in
Figure \ref{fig:energy}, we plot ${\tilde \epsilon}_n$ as a function of ${\tilde
r}$ for the first three disks of the family, with $n = 1$, $2$ and $3$, for
different values of the parameter ${\tilde \alpha}_n = \alpha/a^n$. Then, for
each value of $n$, we take ${\tilde \alpha}_n = 0.5$, $1$, $1.5$, $2$, $2.5$, 
$3$, $3.5$ and $4$. The first curve on left corresponds to ${\tilde \alpha}_n =
0.5$, whereas that the last curve on right corresponds to ${\tilde \alpha}_n =
4$. As we can see, in all the cases the surface energy density is everywhere
positive, having a maximun near the inner edge of the disks, and then rapidly
decreasing as ${\tilde r}$ increases. We can also see that, for a fixed value of
$n$, as the value of ${\tilde \alpha}_n$ increases, the value of the maximum
diminishes and moves towards increasing values of ${\tilde r}$. The same
behavior is observed for a fixed value of ${\tilde \alpha}_n$ and increasing
values of $n$.

On the other hand, by using (\ref{eq:mass}), (\ref{eq:rxy}) and (\ref{eq:mun}),
the total mass of the disks can be expressed as
\begin{equation}
M_n = 2 \pi a \int_0^{\infty} f_n (x) dx,
\label{eq:massn}
\end{equation}
where
\begin{equation}
f_n (x) = x e^{-\Lambda_n (x,0)} \epsilon_n (x). \label{eq:fn}
\end{equation}
So, fom (\ref{eq:lamn}), (\ref{eq:enern}) and (\ref{eq:fn}), is easy to see that
\begin{subequations}\begin{eqnarray}
\lim_{x \to \infty} \frac{f_{n+1} (x)}{f_n (x)} &=& \lim_{x \to \infty}
\frac{\epsilon_{n+1} (x)}{\epsilon_n (x)}, \\
&&	\nonumber	\\
&=& \lim_{x \to
\infty} \frac{E_{n+1} (x)}{a x^2 E_n (x)},	\\
&&	\nonumber	\\
&=& \left\{ \begin{array}{lcl}
\frac{1}{a} &;& n = 2k+1, \\
	&&	\\
0 &;& n = 2k+2, \\
\end{array}\right.
\end{eqnarray}\end{subequations}
with $k \geq 0$. Therefore, by the limit comparison test for improper integrals,
the convergence of $M_{n+1}$ is granted if $M_n$ is convergent and thus we only
need to test the convergence of $M_1$. Indeed, a simple computation gives
\begin{equation}
M_1 = 2 \pi \sqrt{2 a \alpha } \ \Gamma(1/4),
\end{equation} 
by granting so the convergence of all the mass integrals (\ref{eq:massn}).
Accordingly, although the disks are of infinite extension, all of them have
finite mass.

\section{Concluding remarks}\label{sec:con}

We presented an infinite family of asymptotically flat and everywhere regular
exact solutions of the Einstein vacuum equations. These solutions describe an
infinite family of thin dust disks with a central inner edge, whose energy
densities are everywhere positive and well behaved, in such a way that their
energy-momentum tensor are in fully agreement with all the energy conditions.
Moreover, although the disks are of infinite extension, all of them have finite
mass. Now, as all the metric functions of the solutions are explicitly computed,
these are the first fully integrated explicit exact solutions for such kind of
thin disk sources. Furthermore, their relative simplicity when expressed in
terms of oblate spheroidal coordinates, makes it very easy to study different
dynamical aspects, like the motion of particles inside and outside the disks and
the stability of the orbits.

Now, besides their importance as a new family of exact and explicit solutions of
the Einstein vacuum equations, the main importance of this family of solutions
is that they can be easily superposed with the Schwarzschild solution in order
to describe binary systems composed by a thin disk surrounding a central black
hole. Indeed, the superposition of the first member of this family with a
Schwarzschild black hole already has been done, and was previously presented in
\cite{GG1}, whereas that in a subsequent paper a detailed analysis of the
corresponding superposition for the full family will be presented. Accordingly,
like was the family presented here, its superposition with the Schwarzschild
black hole will be the first family of explictly integrated exact solutions for
this superpostion of sources. 

Finally, it is worth to mention an interesting feature of this family of
solutions. As in Newtonian theory the gravitational potential is given by the
solution of the boundary value problem for the Laplace equation, we can consider
the $\Phi_n(r,z;a)$ as a family of Newtonian gravitational potentials of thin
disklike sources with an inner edge, whose Newtonian mass densities are given by
\begin{equation}
\sigma_n (x) = \frac{2 \alpha E_n (x)}{a^{n+1} x^{2n+1}}, \label{eq:newden}
\end{equation}
clearly diverging at the edge of the disks. Accordingly, we can conclude that
there are not regular solutions within the Newtonian theory that properly
describe the gravitational field of a thin disk with an inner edge, whereas that
this kind of source can be properly described, by means of regular and
asymptotically flat solutions, within the general relativistic garvitation
theory.

\begin{acknowledgments}
A. C.~Guti\'errez-Pi\~{n}eres wants to thank the financial support from
COLCIENCIAS, Colombia. The authors would also like to thank S. R. Oliveira for
some valuable comments.
\end{acknowledgments}

\end{document}